\documentclass[12pt]{article}

\usepackage{epsfig,multicol,multirow}
\usepackage{amsmath,subfigure,latexsym,amssymb}

\def\lsi{\raise0.3ex\hbox{$<$\kern-0.75em\raise-1.1ex\hbox{$\sim$}}}
\def\gsi{\raise0.3ex\hbox{$>$\kern-0.75em\raise-1.1ex\hbox{$\sim$}}}
\def\backder{\raise1.4ex\hbox{$\leftarrow$\kern-0.75em\raise-1.4ex\hbox{$\partial$}}}
\def\forder{\raise1.4ex\hbox{$\rightarrow$\kern-0.75em\raise-1.4ex\hbox{$\partial$}}}

\newcommand{\be}{\begin{equation}}
\newcommand{\ee}{\end{equation}}
\newcommand{\nn}{\nonumber}
\newcommand{\bea}{\begin{eqnarray}}
\newcommand{\eea}{\end{eqnarray}}

\newcommand{\uno}{1 \!\! 1}

\newcommand{\R}{{\kern+.25em\sf{R}\kern-.78em\sf{I} \kern+.78em\kern-.25em}}
\newcommand{\RR}{{\kern+.25em\sf{R}\kern-.6em\sf{I} \kern+.6em\kern-.25em}}
\newcommand{\N}{{\kern+.25em\sf{N}\kern-.78em\sf{I} \kern+.78em\kern-.25em}}
\newcommand{\C}{{\kern+.25em\sf{C}\kern-.45em\sf{I} \kern+.45em\kern-.25em}}

\newcommand{\ri}{{\rm i}}

\makeatletter
\@addtoreset{equation}{section}
\makeatother

\begin{document}

\begin{center}

{\Large\bf On the Isomorphic Description} \\
\vspace*{4mm}
{\Large\bf of Chiral Symmetry Breaking} \\
\vspace*{4mm}
{\Large\bf by Non-Unitary Lie Groups} \\
\vspace*{1.1cm}

Wolfgang Bietenholz \\

\vspace*{8mm}

Instituto de Ciencias Nucleares \vspace*{1mm} \\
Universidad Nacional Aut\'{o}noma de M\'{e}xico \vspace*{1mm} \\
A.P. 70-543, C.P. 04510 M\'{e}xico, Distrito Federal \vspace*{2mm} \\
M\'{e}xico \\
\end{center}

\vspace*{0.8cm}

\noindent
It is well-known that chiral symmetry breaking ($\chi$SB) in QCD
with $N_{f}=2$ light quark flavours can be described by orthogonal
groups as $O(4) \to O(3)$, due to local isomorphisms. Here we discuss
the question how specific this property is. We consider generalised 
forms of $\chi$SB involving an arbitrary number of light flavours
of continuum or lattice fermions, in various representations. 
We search systematically for isomorphic descriptions by non-unitary, 
compact Lie groups. It turns out that there are a few alternative 
options in terms of orthogonal groups, while we did not find any 
description entirely based on symplectic or exceptional Lie groups.
If we adapt such an alternative as the symmetry breaking pattern
for a generalised Higgs mechanism, we may consider a Higgs particle 
composed of bound fermions and trace back the mass generation 
to $\chi$SB. In fact, some of the patterns that we encounter 
appear in technicolour models. 
In particular if one observes a Higgs mechanism that
can be expressed in terms of orthogonal groups, we specify in 
which cases it could also represent some 
kind of $\chi$SB of techniquarks.\\

\vspace*{0.6cm}

\noindent
{\footnotesize Keyword: chiral symmetry breaking,
Lie groups, 
Diophantine equations, \\ lattice fermions, technicolour \\

 }


\newpage

\section{Chiral flavour symmetry breaking}

We start by briefly reviewing the process of $\chi$SB
in the physically relevant case of QCD with $N_{f}=2$ quark flavours,
which have masses far below the intrinsic scale 
$\Lambda_{\rm QCD}$.\footnote{This property represents an interesting
hierarchy problem. The existence of this problem is sometimes denied,
based on the argument that light fermions are protected from strong
mass renormalisation by approximate chiral symmetry. In a 
non-perturbative framework, however, it is difficult
to implement (approximate) chiral symmetry. This has been 
achieved in sophisticated ways \cite{KoSus,chilatfer}, but they do still
not make light fermions appear natural. An attempt to arrange for
this in a brane world model is discussed 
in Ref.\ \cite{BGW}.}
In a low energy picture restricted to these two flavours,
the QCD Lagrangian can be written as
\bea
{\cal L} &=& \ri \bar q_{\rm L} {\bf D} q_{\rm L} + 
\ri \bar q_{\rm R} {\bf D} q_{\rm R}
+ m_{\rm u} (\bar u_{\rm L} u_{\rm R} + \bar u_{\rm R} u_{\rm L} ) \nn \\
& & + \ m_{\rm d} (\bar d_{\rm L} d_{\rm R} + \bar d_{\rm R} d_{\rm L})
+  {\cal L}_{\rm pure~gauge} \ .
\eea
In this notation, $u$ and $d$ are spinor fields for the two quark 
flavours, which are decomposed into left- and right-handed components
by the chiral projectors,
\be  \label{chipro}
u_{\rm L,R} = \frac{1}{2} ( \uno \pm \gamma_{5}) u \ , \qquad
\bar u_{\rm L,R} = \frac{1}{2} \bar u ( \uno \mp \gamma_{5}) \quad
{\rm etc.}   
\ee
We further used the short-hand notations
$q_{\rm L} = \left( \begin{array}{c} u_{\rm L} \\ d_{\rm L} \end{array} 
\right)$,
$q_{\rm R} = \left( \begin{array}{c} u_{\rm R} \\ d_{\rm R} \end{array} 
\right)$, 
$\bar q_{\rm L} = ( \bar u_{\rm L}, \bar d_{\rm L})$,
$\bar q_{\rm R} = ( \bar u_{\rm R}, \bar d_{\rm R})$,
and ${\bf D} = \left( \begin{array}{cc} D & \\ & D \end{array} 
\right)$, where $D$ is the (massless) Dirac operator.
In the chiral limit of vanishing quark masses, 
$m_{\rm u}, \, m_{\rm d} \to 0$, the left- and right-handed
components decouple. Therefore $\bar q_{\rm L}, \, q_{\rm L}$
on one hand, and $\bar q_{\rm R}, \, q_{\rm R}$ on the other hand,
can be rotated by arbitrary $U(2)$ transformations, keeping
${\cal L} $ invariant. Thus the Lagrangian has the global symmetry
\be  \label{U2SB}
U(2)_{\rm L} \otimes U(2)_{\rm R} = SU(2)_{\rm L} \otimes SU(2)_{\rm R}
\otimes U(1)_{\rm V} \otimes U(1)_{\rm A} \ .  
\ee
On the right-hand-side of eq.\ (\ref{U2SB}) we split off the vectorial
subgroup $U(1)_{\rm V}$ of simultaneous (L and R) phase rotations,
which is related to the conservation of the baryon number.
The additionally separated subgroup $U(1)_{\rm A}$ for opposite
(L vs.\ R) phase rotations is the axial symmetry, which is explicitly 
broken in QCD through an anomaly.

We focus on the remaining {\em chiral flavour symmetry}, which takes 
for $N_{f}$ massless quark flavours the form
$SU(N_{f})_{\rm L} \otimes SU(N_{f})_{\rm R}$. 
One generally assumes that QCD in the chiral limit (and infinite
volume) would perform spontaneous $\chi$SB,
\be  \label{comprep}
SU(N_{f})_{\rm L} \otimes SU(N_{f})_{\rm R} \to SU(N_{f})_{\rm V} \ ,
\ee
where the vectorial group $SU(N_{f})_{\rm V}$ corresponds again to
simultaneous L and R transformations. According to the Vafa-Witten
Theorem, this remaining flavour symmetry cannot break 
spontaneously \cite{VaWi}. Hence this process yields 
$N_{f}^{2} -1$ Nambu-Goldstone bosons (NGBs).

In Nature the light quarks are not exactly massless; a
small explicit symmetry breaking is superimposed, so that
the NGBs turn into light quasi-NGBs. For $N_{f} = 2$ they are
identified with the pion triplet $\pi^{+}, \, \pi^{0}, \, \pi^{-}$.
If one further includes the (somewhat heavier) $s$-quark, the
quasi-NGBs also embrace the kaons and the $\eta$-particle.
Chiral perturbation theory \cite{XPT} deals with these 
quasi-NGBs as an effective approach to low energy QCD.

This formalism works best for $N_{f}=2$. In this case,
$\chi$SB can be described alternatively as
\be  \label{O4O3}
O(4) \to O(3)
\ee
due to the local isomorphisms $SU(2) \otimes SU(2) \sim O(4)$,
and  $SU(2) \sim O(3)$.
Thus the orthogonal groups provide an equivalent effective picture 
of soft pion physics, which is often more convenient.
Discussions of quasi-spontaneous symmetry breaking $O(N) \to
O(N-1)$, where a weak external ``magnetic field'' provides a small
mass to the $N-1$ quasi-NGBs, can be found for instance in 
Refs.\ \cite{HasLeu}. 

Also studies of {\em generalised forms of $\chi$SB} have a long history, 
for early versions see {\it e.g.}\ Refs.\ \cite{TOW}. A later motivation
--- closer to our work --- departs from the fact that quarks also
interact weakly, hence $\chi$SB in QCD ``breaks''\footnote{We adapt 
here a wide-spread terminology, with inverted commas, however,
because strictly speaking a gauge symmetry can never break, see 
{\it e.g.}\ Ref.\ \cite{suda}.}
the electroweak gauge symmetry and generates a small contribution 
to the $W^{\pm}$- and $Z^{0}$-mass, without involving the Higgs field. 

The concept of {\em technicolour models} is to replace the Higgs 
sector completely by a mechanism of this kind at high energy: 
new fermions {\em (techniquarks)} are added to the Standard Model.
They are confined by a gauge group beyond the Standard Model. At low 
energy they build condensates, which induce $m_{W}, m_{Z}$, while the
hierarchy problem is controlled due to asymptotic freedom
(for reviews, see Refs.\ \cite{technirev}).
In this approach the Higgs particle consists of tightly 
bound fermions, in some analogy to the Cooper pairs in superconductors, 
or even more in superfluids, since the broken symmetry is global.

The Higgs sector of the Standard Model (before gauging)
follows the symmetry breaking pattern (\ref{O4O3}), 
so we have also there the choice between
the use of special unitary or orthogonal groups.
Hence it is indeed tempting 
to try to interpret the Higgs particle as an object composed of tightly 
bound fermions with $N_{f} =2$. 
The question if this works out explicitly is debated in the 
literature, but it is not the concern of this work.
Here we discuss the question if an analogous interpretation is
still conceivable if one observes --- up to moderate energy --- some
Higgs mechanism following a non-standard pattern, involving
compact Lie groups different from the transition (\ref{O4O3}).
Our consideration leads to a list which specifies in which
cases such a pattern is isomorphic to any kind of $\chi$SB.
{\em If} such an isomorphic $\chi$SB process exists, the door
is open for speculations that the Higgs particle is composed of
bound fermions, which might be manifest at very high energy.


The group theoretical properties that we employ
are certainly encoded in the comprehensive mathematical
literature on Lie groups, see {\it e.g.}\ Refs.\ 
\cite{Liebooks}.\footnote{In particular the book by
R.\ Gilmore is useful in the present context.}
This note is physics-oriented and 
focuses on conceivable $\chi$SB processes.

\section{A very general perspective on chirality}

We first adapt a very general perspective, where chirality
just means a global symmetry in the form of two equal but
independent groups $G$, which breaks down to one such
symmetry group. (We are not yet concerned
with corresponding fermion representations.)
Schematically we could write
\be  \label{schema}
G_{\rm L} \otimes G_{\rm R} \to G_{\rm V} \ ,
\ee
but all we really use at this point is the property that the number of
group generators --- the {\em order} of the symmetry group ---
is {\em divided by 2.} In fact this allows for even more
general options than scheme (\ref{schema}). 
We want to check if such a transition could
be described by orthogonal groups, according to
\be  \label{ortho}
\qquad \qquad \qquad \qquad \qquad \qquad \quad
O(N) \to O(n) \qquad \qquad ( N > n ) \ .
\ee 
The group orders imply the condition
$ N (N-1) = 2 n (n-1)$. With the ansatz $k := N-n$ we obtain
\be  \label{Nksq}
N = \frac{1}{2} \left[ 4k + 1 \pm \sqrt{8k^{2} +1} \, \right] \ .
\ee
The argument of the square root must be an odd square number, which
we write as $(2 \ell + 1)^{2}$. This takes the condition to the form
\be  \label{sqtria}
k^{2} = \frac{1}{2}\ell ( \ell + 1) \ .
\ee
So we are looking for numbers, which are doubly figurative,
namely the {\em square triangular numbers} ${\cal F}_{i}$. This is a 
classical problem in number theory \cite{numtheo}. 
There is an infinite string of 
(rapidly growing) solutions, which can be written iteratively as
\be
{\cal F}_{0} = 0 \ , \quad  {\cal F}_{1} = 1 \ , 
\quad  {\cal F}_{i+2} = 34 {\cal F}_{i+1} - {\cal F}_{i} + 2
\quad ({\rm for} ~ i \geq 0) \ . 
\ee
Inserting these numbers into eqs.\ (\ref{sqtria}) and (\ref{Nksq}) 
leads to
\be
\left( \begin{array}{c} N \\ n \end{array} \right) =
\left( \begin{array}{c} 4 \\ 3 \end{array} \right) \ , \
\left( \begin{array}{c} 21 \\ 15 \end{array} \right) \ , \
\left( \begin{array}{c} 120 \\ 85 \end{array} \right) \ \dots
\ee
where the first solution is the physical one that
we mentioned in Section 1.\footnote{The negative sign in eq.\
(\ref{Nksq}) never contributes any sensible solution, since it 
always corresponds to $n \leq 0$.}

\section{Chiral fermions in the complex representation}

Let us now be more specific and consider the case of $\chi$SB
as it occurs in QCD. The quarks are in the complex, fundamental 
representation of the colour gauge group, and $\chi$SB follows 
the pattern anticipated in eq.\ (\ref{comprep}).\footnote{Different 
patterns will be addressed in Sections 5 to 7.} 
It turns out that {\em an isomorphic 
description in the form (\ref{ortho}) has solely the well-known
solution $N_{f}=2, \, N=4, \, n=3$.} \\

To demonstrate this, it is sufficient to compare the order
before $\chi$SB, $2 (N_{f}^{2}-1) = \frac{1}{2} N (N-1)$ \
(with $N_{f} \geq 2$), which means
\be  \label{con1}
N = \frac{1}{2} \left[ 1 + \sqrt{ (4 N_{f})^{2} -15} \, \right] 
\ .
\ee
It is easy to see that the square root is integer only for 
$N_{f}=2$.\footnote{The minimal assumption to single out $N_{f}=2$,
$N=4$, $n=3$, is even more modest: it would have been sufficient to 
start from the ansatz in Section 2 and add the condition that the
rank (cf.\ Table \ref{TabOr}) is also divided by 2 under $\chi$SB, 
as scheme (\ref{schema}) suggests. This leaves transition 
(\ref{O4O3}) as the only solution of the form (\ref{ortho}). 
\label{fnrank}} \\

As an extension we also consider the (hypothetical) case
where $\chi$SB involves the full unitary groups,
$U(N_{f})_{\rm L} \otimes U(N_{f})_{\rm R} \to U(N_{f})_{\rm V}$. 
Now condition (\ref{con1}) is modified to
\be  \label{con2}
N = \frac{1}{2} \left[ 1 + \sqrt{ 1 + (4 N_{f})^{2}} \, \right] \ .
\ee
Here the square root is only integer for the physically
pointless case $N_{f}=0$.\footnote{Lattice simulations
in the ``quenched approximation'' generate configurations based
on the probability weight given by the Euclidean gauge action
alone. Generally the contribution of degenerate
fermion flavours to this weight is given
by the fermion determinant to the power $N_{f}$,
hence the quenched approximation corresponds technically to
$N_{f}=0$. It has been used extensively in lattice QCD
because it speeds up the simulations drastically; in this sense,
or in the limit $m \to \infty$ which renders the fermion determinant
constant, $N_{f}=0$ is not completely academic, though
still not physical.\label{quench}}
So without splitting off the phase factors, we would
not find any solution for a description in the form (\ref{ortho}).

That pattern was originally considered in QCD.
For $N_{f}=2$ it would require a light meson quartet,
where the $\eta$-particle is added to the pion triplet.
For $N_{f}=3$ one would have to add the $\eta'$-particle to
extend the light meson octet to a nonet. However, in both
cases the additional meson is too heavy to fit into the multiplet
(this is a facet of the ``$U(1)$ problem''). 
Therefore that pattern was dismissed in 
favour of the scheme sketched in Section 1.

\section{Involving a product of orthogonal groups}

Of course we can ease the conditions for a description
of $\chi$SB by orthogonal groups if we allow for the ansatz
\be
O(n) \otimes O(n) \to O(n)
\ee
instead of scheme (\ref{ortho}). Then the only condition is a local
isomorphism 
\be  \label{SUO}
SU(N_{f}) \sim O(n) \ .
\ee
Counting once more the generators leads to the Diophantine equation
\be  \label{con3}
N_{f}^{2} -1 = \frac{1}{2} n (n-1) \ ,
\ee
which is not as simple as the cases that we encountered in Section 3.
The general formula for inductive solutions leads to
\be
\left( \begin{array}{c} N_{f}^{(i+1)} \\ n^{(i+1)} \end{array} \right)
= \left( \begin{array}{cc} 3 & 2 \\ 4 & 3 \end{array} \right) \
\left( \begin{array}{c} N_{f}^{(i)} \\ n^{(i)} \end{array} \right) 
- \left( \begin{array}{c} 1 \\ 1 \end{array} \right)
\ .
\ee
It can be obtained conveniently from D.\ Alpern's Online Calculator
\cite{Dario}, and Refs.\ \cite{numtheo} review its number
theoretical background.
We arrive at solutions with $N_{f} > 1$ by starting from
$\left( \begin{array}{c} N_{f}^{(0)} \\ n^{(0)} \end{array} \right)
= \left( \begin{array}{c} 1 \\ 0 \end{array} \right)$ or
$\left( \begin{array}{c} 1 \\ 1 \end{array} \right)$.
This yields two strings of solutions,
\bea
\left( \begin{array}{c} N_{f} \\ n \end{array} \right) &=&
\left( \begin{array}{c} 2 \\ 3 \end{array} \right) \ , \
\left( \begin{array}{c} 11 \\ 16 \end{array} \right) \ , \
\left( \begin{array}{c} 64 \\ 91 \end{array} \right) \ \dots \nn \\
&=&
\left( \begin{array}{c} 4 \\ 6 \end{array} \right) \ , \
\left( \begin{array}{c} 23 \\ 33 \end{array} \right) \ , \
\left( \begin{array}{c} 134 \\ 190 \end{array} \right) \ \dots 
\eea

However, so far we have only considered the necessary condition
for the orders to matches. Of course, an isomorphism 
requires more than that. Now that we have a set of solution
candidates, we compare as a further criterion the {\em rank},
{\it i.e.}\ the number of simultaneously diagonalisable
generators. It amounts to $N_{f}-1$ for $SU(N_{f})$, and for
$O(n)$ it is $[n/2]$, which means $n/2$ if $n$ is even, 
and $(n-1)/2$ if $n$ is odd, cf.\ Table \ref{TabOr}.

We combine this condition with eq.\ (\ref{con3}), eliminate
$n$ and solve for $N_{f}$. This yields two solutions 
with $N_{f}>1$,
\be
N_{f} = 2 \ , \ n = 3 \qquad {\rm or}  \qquad
N_{f} = 4 \ , \ n = 6 \ .
\ee
Indeed these are the two cases where the isomorphism
(\ref{SUO}) is known to work \cite{Liebooks}. The former case is 
once more equivalent to the solution anticipated in Section 1,
if we add $O(3) \otimes O(3) \sim O(4)$. The second case,
\be  \label{SU4}
SU(4)_{\rm L} \otimes SU(4)_{\rm R} \to SU(4)_{\rm V} \quad \sim 
\quad O(6)_{\rm L} \otimes O(6)_{\rm R} \to O(6)_{\rm V}
\ee
can be viewed as the only alternative description of $\chi$SB
in the complex representation in terms of orthogonal groups.
In QCD it would mean to include even the  $c$-quark into the
 $\chi$SB scheme. However, its mass of $m_{\rm c} \simeq 1.3 ~ 
{\rm GeV}$ is too heavy to be captured by chiral perturbation theory.

\section{$\chi$SB in the real or pseudo-real representation}

The literature refers additionally to another two forms of $\chi$SB, 
which we have not covered yet. Studies of technicolour models
pointed out that chiral fermions in four dimensions, 
interaction through a Yang-Mills gauge field, can perform 
exactly three types of spontaneous $\chi$SB,
depending on the representation of the fermion field \cite{Dimo,Peskin}.
In this work we also consider further variants, 
which may occur in explicit $\chi$SB 
(through an asymmetric term in the Lagrangian, like an explicit fermion 
mass in a vector theory), or through an anomaly (as in the 1-flavour
Schwinger model)\footnote{Generally
the chiral condensate 
$\Sigma = - \ ^{\lim}_{m \to 0} \ ^{\lim}_{V \to \infty}
\langle \bar \Psi \Psi \rangle$ is the order parameter for 
spontaneous $\chi$SB, where $\bar \Psi$,  $\Psi$ incorporate
all fermion components ($m$ is the fermion mass and $V$ the volume).
In the chiral limit of the Schwinger 
model (2d QED) it is ill-defined in the quenched case ($N_{f}=0$, cf.\
footnote \ref{quench}) \cite{Dam2}, 
but it takes a finite value for $N_{f}=1$ \cite{Cole} (here chiral
symmetry simply means invariance under 
$\psi \to \exp (i \alpha \gamma_{5}) \psi$, 
$\bar \psi \to \bar \psi \exp (i \alpha \gamma_{5})$ ($\alpha \in \RR$)). 
There the $\chi$SB pattern agrees with QCD. 
For $N_{f} > 1$, $\Sigma$ vanishes; at finite $m$ it is an 
example for $\chi$SB not matching any of the three patterns established 
in Refs.\ \cite{Dimo,Peskin}. This can be seen from the microscopic 
Dirac spectrum \cite{Lat07}, since no Banks-Casher plateau \cite{BC}
emerges (cf.\ subsequent remarks on Random Matrix Theory).}
or through the regularisation
(as in the case of staggered lattice fermions, see Section 7).

The current section, however, does address the three $\chi$SB
patterns that Refs.\ \cite{Dimo,Peskin} referred to.
To present them, we consider the
Dirac matrices in the Weyl representation, where the chiral
projectors of eq.\ (\ref{chipro}) are diagonal. 
Then the Dirac operator (in a Yang-Mills gauge
background) has a off-diagonal block structure, which takes
in Euclidean space the form
\be
D = \gamma_{\mu} \left( \ri \partial_{\mu} + g A_{\mu} \right)
= \left( \begin{array}{cc} 0 & d \\ d^{\dagger} & 0 \end{array}
\right) \ , \qquad
{\rm with}~~ A_{\mu} =  \sum_{a=1}^{N_{c}} A_{\mu}^{a} T_{a} \ ,
\ee
where $T_{a}$ are the generators of the gauge group, and $g$
is the gauge coupling.

We have considered so far the case $N_{c} \geq 3$ and $A_{\mu}$
in the fundamental representation, with matrix elements
$d_{ij} \in \C$. In this case the irreducible Dirac fermion 
representation of the gauge group is complex
{\em (complex representation)}. One alternative is
the case $N_{c} =2$, still in the fundamental representation,
where an additional symmetry ensures $d_{ij} \in \R$
{\em (real representation).}
Finally, for $N_{c} \geq 3$ but $A_{\mu}$ in the adjoint
representation, the matrix elements $d_{ij}$ are real quaternionic
{\em (pseudo-real representation)}, as summarised 
in Refs.\ \cite{Poul}. Ref.\ \cite{Jacques} discussed
the Dirac spectra in these three classes. They match the spectra
in three distinct types of Random Matrices, which had been identified by
F.J.\ Dyson \cite{Dyson}. Numerical simulations with chiral
lattice quarks \cite{chilatfer} (without doubling) confirm that QCD 
obeys the predictions for the complex representation \cite{WBStani};
this also captures the distinction between the Random Matrix
formulae for different topological sectors \cite{RMTeq}. \\

In the other two cases occurring in 4d Yang-Mills theory,
quark and anti-quark representations are equivalent, hence
the (unbroken) chiral symmetry group is enlarged to 
$SU(2 N_{f})$, and the $\chi$SB patterns are 
\bea
{\rm real~:} & SU(2 N_{f}) \to SO(2 N_{f}) &  \qquad
2 N_{f}^{2} +  N_{f} -1 ~~ {\rm NGBs} \nn \\
\mbox{pseudo-real~:} & SU(2 N_{f}) \to Sp(2 N_{f}) & \qquad
2 N_{f}^{2} -  N_{f} -1 ~~ {\rm NGBs} \ . \nn
\eea
(The ansatz in Section 2 was a broad generalisation, but still 
restricted to an (extended) framework of the complex representation.)

These schemes can be illustrated \cite{Peskin} by writing 
the fermion fields as vectors consisting of $2 N_{f}$ 
(2-component) spinors,
$\Psi = \left( \begin{array}{c} \psi_{1} \\ \psi_{2} \end{array} \right)$,
$\Phi = \left( \begin{array}{c} \phi_{1} \\ \phi_{2} \end{array} \right)$.
$\Psi$ and $\Phi$ are in the fundamental representation of $SU(2N_{f})$
(the anomalous phase being split off), and $\psi_{i}, \ \phi_{i}$
are composed of $N_{f}$ spinors.

With the definition $\psi^{\pm} = \frac{1}{\sqrt{2}} (\psi_{1} \pm
{\rm i} \psi_{2})$ (and $\phi^{\pm}$ analogous) the scalar product
can be written as
$
\Psi^{\dagger} \chi = (\psi^{+}, \psi^{-})^{\dagger} 
\left( \begin{array}{c} \phi^{+} \\ \phi^{-} \end{array} \right) ,
$
which shows its invariance under the transformations
\be
\begin{array}{c} \psi^{+} \to U \psi^{+} \\
\psi^{-} \to U^{*} \psi^{-} \end{array} \quad U \in U(N_{f})
\quad {\rm and} \quad
\begin{array}{c} \psi^{+} \to V \psi^{+} \\
\psi^{-} \to V^{T} \psi^{-} \end{array} \quad V \in SU(N_{f}) 
\label{trafoUV}
\ee
(and the same for $\phi^{\pm}$). Together they build the
$SU(N_{f}) \otimes SU(N_{f}) \otimes U(1)$ subgroup of $SU(2N_{f})$,
which is relevant for $\chi$SB in the complex representation.

To capture the other two options, note that the following
bilinear forms are invariant under further subgroups of $SU(2N_{f})$,
\begin{eqnarray*}
s & \hspace*{-2.4cm}
= \Psi^{T} \Phi & {\rm preserved~under} \quad O(2N_{f}) \\
a & = \Psi^{T} \left( \begin{array}{cc} 0 & \uno \\ -\uno & 0
\end{array} \right) \Phi & {\rm preserved~under} \quad Sp(2N_{f}) \ .
\end{eqnarray*}
The properties $2 (\psi^{+})^{T} \phi^{-} = s - {\rm i} a$, \
$2 (\psi^{-})^{T} \phi^{+} = s + {\rm i} a$, show that
the transformation matrix $U$ in (\ref{trafoUV}) preserves $s$ and 
$a$, hence $U(N_{f})$ is a subgroup of both, $O(2N_{f})$
and $Sp(2N_{f})$ 
\cite{Liebooks}. Now the generators for the various subgroups of
$SU(2N_{f})$ can be extracted \cite{Peskin}, confirming the
orders displayed in Table \ref{TabOr}. \\

We recall that $Sp(2N)$ is the group of symplectic transformations, which 
can be represented by real $2N \times 2N$ matrices \cite{Liebooks}. 
We have used above the property that $Sp(2N)$ has order 
$N ( 2N +1)$, and its rank is $N$, as indicated in Table \ref{TabOr}. 
Counting the number of NGBs, we note that neither of
these options --- real or pseudo-real --- could explain the pion triplet,
or the light meson octet, which are observed in Nature. This confirms
once more that only the complex representation is relevant for the
strong interaction.

Let us nevertheless check the possibilities of a description
by orthogonal groups also for these additional types of $\chi$SB:
\begin{itemize}
\item In the {\em real} representation the issue is only to find 
an isomorphism of $SU(2N_{f})$ to some orthogonal group $O(N)$.
We saw in Section 4 that this works out in two
cases: $N_{f}=1, \ n=3$, or $N_{f}=2, \ n=6$.
\item In the {\em pseudo-real} representation we have to match in 
addition $Sp(2N_{f})$ with some group $O(n)$ (or $SO(n)$). If we set 
order and rank equal, we see that $n$ must be odd, and we are left with
two solutions: $N_{f}=1, \ n=3$ and $N_{f}=2, \ n=5$.

In the first case no $\chi$SB takes place, since 
$SU(2) \sim Sp(2) \sim O(3)$, in agreement with the
vanishing number of NGBs being generated.

Regarding the second case, the isomorphism $Sp(4) \sim O(5)$
does in fact hold \cite{Liebooks}, so the pseudo-real $\chi$SB 
for $N_{f}=2$ can be described as $O(6) \to O(5)$. This transition
is also covered by Refs.\ \cite{HasLeu}.
\end{itemize}

\section{Probing symplectic and exceptional \\ Lie groups}

Next we address the question if $\chi$SB could be described
isomorphically in terms of symplectic or exceptional Lie groups,
as an alternative to the orthogonal groups that we have considered
so far. Note that these alternatives are compact as well.
For the very general perspective of Section 2, the
ansatz $Sp(2N) \to Sp(2n)$ leads again to a non-trivial
Diophantine equation, $N(2N+1) = 2n (2n+1)$, with the recursive
solution \cite{numtheo,Dario}
\be
\left( \begin{array}{c} N^{(i+1)} \\ n^{(i+1)} \end{array} \right)
= \left( \begin{array}{cc} 17 & 24 \\ 12 & 17 \end{array} \right)
\left( \begin{array}{c} N^{(i)} \\ n^{(i)} \end{array} \right) +
\left( \begin{array}{c} 10 \\ 7 \end{array} \right) \ .
\ee
We may start from the trivial solution $N=n=0$, which yields the 
sequence\footnote{If we require also the rank to be divided by 2, 
as in footnote \ref{fnrank}, no non-trivial solution persists.}
\be
\left( \begin{array}{c} N \\ n \end{array} \right)  
= 
\left( \begin{array}{c} 10 \\ 7 \end{array} \right), \
\left( \begin{array}{c} 348 \\ 246 \end{array} \right) \ \dots
\ee

Let us now include the exceptional Lie groups $G_{2}, \, 
F_{4}, \, E_{6}, \, E_{7}$ and $E_{8}$ into the consideration
(see {\it e.g.}\ Ref.\ \cite{except}), 
and we also allow for transitions between different types of groups.
As long as we solely require the order to be divided by $2$,
we find further options, such as $Sp(10) \to O(15)$, $O(21) \to
Sp(14)$ or $O(32) \to E_{8}$. Moreover, in the solution
$O(8) \to G_{2}$ also the rank is divided by $2$. However, it does
still not match the pattern (\ref{schema}), and in none of these cases
the unbroken symmetry can be identified with some kind of chiral
symmetry in the usual sense. \\

For the options that occurred in Sections 3 to 5, such an identification
with chiral symmetry holds. However, in all these cases
we would need some isomorphism of the group type that we focus on 
to $SU(N_{f})$. For the latter the order $\Omega$ and the rank
$r$ are related as 
\be  \label{Or}
\Omega = r (r+2) \ .
\ee
For the symmetry groups under consideration here, we display
$\Omega$ and $r$ in Table \ref{TabOr}.
\begin{table}
\centering   
\begin{tabular}{|c||c|c|c|c|c|c|c|c|}
\hline
Group & $SU(N)$ & $O(N)$ & $Sp(2N)$ &  $G_{2}$ & $F_{4}$ & $E_{6}$ 
& $E_{7}$ & $E_{8}$ \\
\hline
\hline
 order $\Omega$ & $N^{2}-1$ & $\frac{1}{2} N (N-1)$ &
$N(2N+1)$ & $14$ & $52$ & $78$ & $133$ & $248$ \\ 
\hline 
 rank $r$ & $N-1$ & $[N/2]$ & $N$ & $2$ & $4$ & $6$ & $7$ & $8$ \\  
\hline
\end{tabular}
\caption{The order and rank of various Lie groups that
we considered. ($[N/2]$ means the integer among $N/2$ and $(N-1)/2$.)}
\label{TabOr}
\end{table}
We see that the relation $Sp(2) \sim SU(2)$, which we encountered
before in Section 5, is the only solution to eq.\ (\ref{Or})
among the symplectic or exceptional groups. This
relation is actually an identity \cite{Liebooks}, so one should 
not regard it as an alternative description. \\

If we reconsider unitary (instead of special unitary) $\chi$SB,
relation (\ref{Or}) turns into $\Omega = r^{2}$, which cannot
be matched by any symplectic or exceptional group, see Table \ref{TabOr}.

If we also take $SU(N)$ and $O(N)$ into account (and exclude
the trivial group $SU(1)$), the only group of Table \ref{TabOr} obeying
$\Omega = r^{2}$ is $O(2)$. However, we will not include the transition
$U(1)_{\rm L} \otimes U(1)_{\rm R} \to U(1)_{\rm V} \ \sim \
O(2)_{\rm L} \otimes O(2)_{\rm R} \to O(2)_{\rm V}$ in our concluding 
list in Section 8, because it does not agree with the usual notion 
of $\chi$SB (in the continuum).

\section{$\chi$SB for lattice fermions}

Let us finally address further $\chi$SB patterns, which occur
in non-perturbative studies by means of numerical simulations 
of vector theories on the lattice. 
Traditionally two types of lattice fermion formulations 
have usually been applied in Monte Carlo simulations.
In one of them, the Wilson fermion \cite{Wilfer}, a discrete 
Laplacian term is added to the naive discretisation in order to
avoid the fermion doubling. This term breaks the chiral symmetry
explicitly, and the chiral limit can only be attained by
fine-tuning the bare fermion mass. The pattern of $\chi$SB,
however, for Wilson fermions (and variants thereof) is
the same as in the continuum. 

The same holds for Ginsparg-Wilson fermions \cite{chilatfer}, 
which gained importance only recently:
they obey an exact, lattice modified chiral symmetry \cite{ML},
which turns into the standard chiral symmetry in the continuum
limit. This modified chiral symmetry prevents additive mass 
renormalisation, so the chiral limit does not require fine tuning,
and also at finite lattice spacing the flavour chiral symmetry
breaking follows the same pattern as in the continuum (although
the symmetry transformation is local in this case). \\

The situation is {\em different,} however, for the second
traditional standard lattice fermion, denoted as the 
{\em staggered fermion} \cite{KoSus}. 
Unlike the Wilson fermion it does not suffer from additive mass 
renormalisation. 
In this respect it is an alternative to the Ginsparg-Wilson fermion; 
the staggered fermion is much simpler to simulate, but
plagued with unpleasant constraints on $N_{f}$ (if one insists on
locality in order to assure a controlled continuum limit \cite{Creutz}).

In its construction one starts from the naive lattice fermion, which
is doubled in each direction. By means of a lattice site dependent
transformation, the $\gamma$-matrix structure can be removed,
so that one only needs to keep track of 1 out of the original 4
spinor components (for $N_{f}=1$ in 4 dimensions). One distributes 
its $16$ copies over the sites of unit hypercubes on the lattice. At 
this point, one distinguishes an {\em even} and an {\em odd} sub-lattice 
(it consists of the sites where the sum of the coordinates in lattice 
units, $x_{1} + x_{2}+x_{3}+x_{4}$, is even resp.\ odd). In the chiral
limit, the staggered fermion components on these two sub-lattices can 
be rotated independently in $\C$ without altering the action, which 
amounts to a global $U(1)_{\rm e} \otimes U(1)_{\rm o}$ symmetry. It 
contains the axial $U(1)$ symmetry, along with a $U(1)$ remnant chiral 
symmetry, whereas the vectorial $U(1)$ group is redundant in this
formulation (see {\it e.g.}\ Ref.\ \cite{Smitbook}). 

In the continuum limit 4 flavours can be assembled, and the lattice
$U(1)$ invariance is remnant of the corresponding $SU(4) \otimes SU(4)$
chiral flavour symmetry. This would again 
correspond to the (inappropriate) inclusion of the $c$-quark, 
as in eq.\ (\ref{SU4}). At finite lattice spacing, the transition
$U(1) \sim O(2) \to 1$ is the numerically observed $\chi$SB.
Hence the simple property referred to in the last paragraph of
Section 6 has some kind of application on the lattice.

For $N_{f}$ staggered fermions in the complex representation,
the global symmetry is extended to 
$U(N_{f})_{\rm e} \otimes U(N_{f})_{\rm o}$.
This is the setting that we addressed in the last paragraphs
of Sections 3 and 6. Now the $\chi$SB pattern yields the coset space
\be
SU(N_{f}) \otimes SU(N_{f}) \otimes U(1) / SU(N_{f}) =  U(N_{f}) \ .
\ee 
This does not allow for any alternative description by non-unitary 
Lie groups (without building direct products), 
except for the $N_{f} =1$ case that we mentioned before.

In the real or pseudo-real representation the chiral symmetry group
is enlarged to $U(2N_{f})$, similar to the $\chi$SB patterns of Section 5.
However, compared to the continuum situation that we addressed
before, for the staggered fermions the non-breaking symmetry is 
interchanged, {\it i.e.}\ the coset space reads $U(2N_{f}) / Sp(2N_{f})$ 
\ ($U(2N_{f}) / SO(2N_{f})$) in the real (pseudo-real) representation 
\cite{Poul}. Here the search for an isomorphic descriptions 
by non-unitary Lie groups (again without direct products)
fails because there is no isomorphism at all to $U(2N_{f})$. \\

At last we mention that simulations are also possible in a Hamiltonian
formulation, though this approach is tedious and therefore not
popular. At strong coupling it leads to the $\chi$SB pattern 
\cite{Smitbook}
\be
U(4 N_{f}) \to U(2 N_{f}) \otimes U(2 N_{f}) \ .
\ee
In this case the order is divided by $2$ (which is the
minimal requirement that we postulated in Section 2), but the
rank remains unchanged. The second property does not hold for any
of the non-unitary isomorphic descriptions that we found to
obey the first property, hence there is no alternative
description for that type of $\chi$SB. (The invariance of the 
rank and the conclusion still persists if we split off an axial 
phase symmetry.)

\section{Conclusions}
\vspace*{-1mm}

The well-known description of Chiral Symmetry Breaking ($\chi$SB)
in $N_{f} =2$ QCD by means of orthogonal groups is
indeed quite specific. We have considered broad
generalisations of $\chi$SB and studied the question if there
are further isomorphic descriptions by non-unitary Lie groups.
We only found a few possibilities in terms of orthogonal groups,
but none with symplectic or exceptional Lie groups. The 
alternatives are summarised in Table \ref{conclutab}.

\begin{table}
\centering   
\begin{tabular}{|c|c||c|c|}
\hline
fermion & $N_{f}$ & $\chi$SB pattern & isomorphic description \\
representation &  &  & without unitary groups \\
\hline
complex & 4 & $SU(4) \otimes SU(4) \to SU(4)$ & 
$O(6) \otimes O(6) \to O(6)$ \\
\hline
real & 1 & $SU(2) \to SO(2)$ & $O(3) \to SO(2)$ \\
\hline
real & 2 & $SU(4) \to SO(4)$ & $O(6) \to SO(4)$ \\
\hline
pseudo-real & 2 & $SU(4) \to Sp(4)$ & $O(6) \to O(5)$ \\
\hline
\end{tabular}
\caption{The list of the $\chi$SB breaking patterns, which are 
isomorphic to some transition that does not involve unitary groups. 
We have demonstrated that this list is in fact {\em complete}.}
\label{conclutab}
\end{table}


In addition the simple $\chi$SB $U(1) \sim O(2) \to 1$
occurs for $N_{f}=1$ staggered fermion at finite
lattice spacing ({\it i.e.}\ on the regularised level).

We mentioned before that there are no applications of the
transitions in Table \ref{conclutab} to QCD with 
the quark masses observed in Nature. Possibly
conceivable applications of the right-most-side of Table 
\ref{conclutab} could be non-standard variants of the
Higgs mechanism (cf.\ Section 1). 
However, the case $O(3) \to SO(2)$ does not provide
a sufficient number of NGBs to generate massive gauge bosons
$W^{\pm}, \, Z^{0}$. In the other cases there would be an 
abundance of $2 \dots 12$ NGBs, and one would have to explain 
why they have not been manifest in low energy phenomenology.
This suggests that additional symmetry breaking would 
be required to render these particles heavy.

Proceeding to the third column of Table \ref{conclutab}, we
are led to scenarios without a fundamental scalar particle,
where the particle masses are generated by the chiral
condensate of (non-standard) fermions. This takes us back to the
technicolour models \cite{technirev} that we already referred 
to in Section 1. Early versions were based on a direct
product of the Standard Model gauge groups with a new 
(technicolour) gauge group,
but since the standard fermions are then technigauge singlets 
they cannot become massive. 

Extended versions ({\it e.g.}\ in Refs.\ \cite{Dimo,Peskin}) 
use a {\em unified gauge group,} which
contains those of the Standard Model (that it breaks down to),
thus coupling techniquarks to standard fermions. This concept
matches our discussion, and fermion masses can be generated,
though the heavy top quark is still uneasy to explain,
and flavour-changing neutral currents emerge, which are not
observed. Minimal Walking Models \cite{MWM} are versions, 
which are fashion now, and which try to avoid this problem
by a near-conformal gauge dynamics so that the coupling
``walks'' instead of ``running''. There are numerous recent
attempts to identify a suitable theory with a ``walking'' 
gauge coupling by means of lattice simulations
\cite{Simon}. 

The issue in these studies is the search for an adequate strongly
interacting model with a conformal window, {\it i.e.}\
with an IR fixed point. Indicators of this property could be
that the ratio $m_{\rm pseudoscalar}/m_{\rm fermion} \propto$
chiral condensate, the string tension and the pseudoscalar decay 
constant vanish in the chiral limit, in contrast to QCD.
So far, the numerical studies suffer from difficulties to
attain the chiral regime.\\

Let us finally summarise the scenario that we mentioned before. 
We assume phenomenology at moderate energy to be consistent with
a (possibly non-standard) Higgs mechanism in terms of orthogonal 
groups. This suggests a multi-component scalar Higgs field. 
 
Now we wonder if the corresponding Higgs particle could still
have a fermionic substructure, which may be manifest at very
high energy (and which could help for instance to overcome the
hierarchy problem). If the orthogonal symmetry breaking pattern
can be identified isomorphically with some kind of $\chi$SB,
this scenario is conceivable. Table \ref{conclutab} presents a
list of the patterns where this is the case. 
If, on the other hand, we observe some orthogonal pattern that
is not included in this list --- or if we observe a pattern 
based on other Lie groups, which do not match any kind of $\chi$SB 
--- such an interpretation is hard to advocate. Hence 
Table \ref{conclutab} distinguishes whether or not an obvious 
techniquark candidate exists. If this is the case, the actual 
viability of such an underlying description is still to 
be investigated. \\




\noindent
{\em Acknowledgement} : I thank Kieran Holland for helpful remarks.


\begin{thebibliography}{10}

\bibitem{KoSus} J.B.\ Kogut and L.\ Susskind, {\it Phys.\ Rev.} 
{\bf D 11} (1975) 395.

\bibitem{chilatfer} P.H.\ Ginsparg and K.G.\ Wilson,
{\it Phys. Rev.} {\bf D 25} (1982) 2649.
D.B.\ Kaplan, {\em Phys.\ Lett.} {\bf B 288} (1992) 342.
W.\ Bietenholz and U.-J.\ Wiese, {\it Nucl.\ Phys.} {\bf B 464} 
(1996) 319. H.\ Neuberger, {\it Phys.\ Lett.} {\bf B 417} (1998) 141.
P.\ Hasenfratz, {\em Nucl.\ Phys.\ (Proc.\ Suppl.)}
{\bf B 63} (1998) 53.

\bibitem{BGW} W.\ Bietenholz, A.\ Gfeller and U.-J.\ Wiese,
{\em JHEP} {\bf 0310} (2003) 018.

\bibitem{VaWi} C.\ Vafa and E.\ Witten,
{\em Nucl.\ Phys.} {\bf B 234} (1984) 173.

\bibitem{XPT} S.\ Weinberg, {\it Physica} {\bf A 96} (1979) 327.
J.\ Gasser and H.\ Leutwyler, {\it Annals Phys.} {\bf 158} (1984) 
142; {\it Phys.\ Lett.} {\bf B 184} (1987) 83.

\bibitem{HasLeu} P.\ Hasenfratz and H.\ Leutwyler, {\it Nucl.\ Phys.}
{\bf B 343} (1990) 241.
W.\ Bietenholz, {\it Helv.\ Phys.\ Acta} {\bf 66} (1993) 633.

\bibitem{TOW} L.\ Turner and M.G.\ Olssen, 
{\it Phys. Rev.} {\bf D 6} (1972) 3522.
D.L.\ Weaver, {\it Annals Phys.} {\bf 101} (1976) 52.

\bibitem{suda} A.\ Perez and D.\ Sudarsky,
arXiv:0811.3181 [hep-th].

\bibitem{technirev} C.T.\ Hill and E.H.\ Simmons,
{\em Phys.\ Rept.} {\bf 381} (2003) 235 
[erratum-ibid.\ {\bf 390} (2004) 553].
F.\ Sannino,
arXiv:0804.0182 [hep-ph].

\bibitem{Liebooks} R.\ Gilmore, ``Lie Groups, Lie Algebras,
and Some of Their Applications'', John Wiley \& Sons (1974).
R.W.\ Carter, ``Simple Groups of the Lie Type'',
John Wiley \& Sons (1989). D.\ Bump, ``Lie Groups'', Springer (2004).


\bibitem{numtheo} 
G.\ Everest and T.\ Ward,
``An introduction to number theory'', Springer (2005).
Y.I.\ Manin and A.A.\ Panchishkin,
``Introduction to modern number theory: 
fundamental problems, ideas and theories'', Springer (2005).
H.\ Cohen, ``Number Theory, Volume I: Tools and Diophantine 
Equations'', Springer (2007).

\bibitem{Dario} D.\ Alpern, Online Calculator, \\
http://www.alpertron.com.ar/CUAD.HTM

\bibitem{Dimo} S.\ Dimopoulos, {\em Nucl.\ Phys.} {\bf B 168} (1980) 69.
J.P.\ Preskill, {\em Nucl.\ Phys.} {\bf B 177} (1981) 21.

\bibitem{Peskin} M.E.\ Peskin, {\em Nucl.\ Phys.} {\bf B 175} (1980) 197. 

\bibitem{Dam2} P.H.\ Damgaard, U.M.\ Heller, R.\ Narayanan
and B.\ Svetitsky,
{\em Phys.\ Rev.} {\bf D 71} (2005) 114503. 

\bibitem{Cole} S.R.\ Coleman, R.\ Jackiw and L.\ Susskind,
{\em Annals Phys.} {\bf 93} (1975) 267.

\bibitem{Lat07} W.\ Bietenholz, S.\ Shcheredin and J.\ Volkholz,
{\em PoS(LAT2007)064}. 
W.\ Bietenholz and I.\ Hip, arXiv:0909.2241 [hep-lat].

\bibitem{BC} T.\ Banks and A.\ Casher,
{\em Nucl.\ Phys.} {\bf B 169} (1980) 103.

\bibitem{Poul} P.H.\ Damgaard, U.M.\ Heller, R.\ Niclasen
and B.\ Svetitsky, {\em Nucl.\ Phys.} {\bf B 633} (2002) 97.
P.H.\ Damgaard, {\it Nucl.\ Phys.\ (Proc.\ Suppl.)} 
{\bf 106} (2002) 29.

\bibitem{Jacques} J.J.M.\ Verbaarschot, {\it Phys.\ Rev.\ Lett.} 
{\bf 72} (1994) 2531. 

\bibitem{Dyson} F.J.\ Dyson, {\it J.\ Math.\ Phys.} {\bf 3} (1962) 140; 
{\it J.\ Math.\ Phys.} {\bf 6} (1962) 1199.

\bibitem{WBStani} W.\ Bietenholz, K.\ Jansen and S.\ Shcheredin,
{\em JHEP} {\bf 07} (2003) 033.
L.~Giusti, M.~L\"uscher, P.~Weisz and H.~Wittig,
\emph{JHEP} {\bf 11} (2003) 023.
D.~Galletly {\it et al.}\ (QCDSF-UKQCD Collaboration),
\emph{Nucl.\ Phys.\ (Proc.\ Suppl.)} {\bf B 129\&130} (2004) 456.

\bibitem{RMTeq} T.\ Wilke, T.\ Guhr and T.\ Wettig,
{\em Phys.\ Rev.} {\bf D 57} (1998) 6486.
P.H.\ Damgaard and S.M.\ Nishigaki,
{\it Phys. Rev.} {\bf D 63} (2001) 045012.

\bibitem{except} J.F.\ Adams, ``Lectures on Exceptional Lie Groups''
(eds.\ Z.\ Mahmud and M.\ Mimura), The University of Chicago Press (1996).

\bibitem{Wilfer} K.G.\ Wilson, in ``New Phenomena in Subnuclear 
Physics'' (ed.\ A.\ Zichichi), Plenum (1979) p.\ 69.

\bibitem{ML} M.\ L\"{u}scher,
\emph{Phys.\ Lett.} {\bf B 428} (1998) 342. 

\bibitem{Creutz} M.\ Creutz,
\emph{Phys.\ Lett.} {\bf B 649} (2007) 230. 

\bibitem{Smitbook} J.\ Smit, ``Introduction to Quantum Fields on a 
Lattice'', Cambridge University Press (2002).

\bibitem{MWM} B.\ Holdom, {\it Phys. Rev.} {\bf D 24} (1981) 1441. 

\bibitem{Simon} S.\ Catterall and F.\ Sannino,
{\it Phys. Rev.} {\bf D 76} (2007) 034504. 
T.\ DeGrand, Y.\ Shamir and B.\ Svetitsky,
{\it Phys. Rev.} {\bf D 78} (2008) 031502;
{\it Phys. Rev.} {\bf D 79} (2009) 034501.
L.\ Del Debbio, A.\ Patella and C.\ Pica,
arXiv:0805.2058 [hep-lat].
S.\ Catterall, J.\ Giedt, F.\ Sannino and J.\ Schneible, 
{\em JHEP} {\bf 0811} (2008) 009.
A.J.\ Hietanen, J.\ Rantaharju, K.\ Rummukainen and K.\ Tuominen, 
{\em JHEP} {\bf 0905} (2009) 025.
A. Deuzeman, M.P.\ Lombardo and E.\ Pallante,
arXiv:0904.4662 [hep-ph].
Z.\ Fodor, K.\ Holland, J.\ Kuti, D.\ Nogradi and C.\ Schroeder,
{\it Phys. Lett.} {\bf B 681} (2009) 353;
arXiv:0908.2466 [hep-lat].
D.K.\ Sinclair and J.B.\ Kogut,
arXiv:0909.2019 [hep-lat].
C.\ Pica, L.\ Del Debbio, B.\ Lucini, A.\ Patella and A.\ Rago,
arXiv:0909.3178 [hep-lat].


\end{thebibliography}
\end{document}